\documentclass{article}
\usepackage{arxiv}

\usepackage{times}
\usepackage{soul}
\usepackage{url}
\usepackage[hidelinks]{hyperref}
\usepackage[utf8]{inputenc}
\usepackage[small]{caption}
\usepackage{subcaption}
\usepackage{graphicx}
\usepackage{amsmath}
\usepackage{amsthm}
\usepackage{booktabs}
\usepackage{algorithm}
\usepackage{algorithmicx}
\usepackage[noEnd=true]{algpseudocodex}
\algrenewcommand\algorithmicindent{1.0em}%
\usepackage[switch]{lineno}
\usepackage{multirow}
\usepackage{rotating}
\usepackage{adjustbox}
\usepackage{tabularx}

\title{Anytime Multi-Agent Path Finding using Operation Parallelism\\in Large Neighborhood Search}

\author{
Shao-Hung Chan$^1$
\and
Zhe Chen$^2$\and
Dian-Lun Lin$^3$\and
Yue Zhang$^2$\and
Daniel Harabor$^2$\and\\
Tsung-Wei Huang$^3$\and
Sven Koenig$^1$\And
Thomy Phan$^1$\\
\affiliations
$^1$Department of Computer Science, University of Southern California, Los Angeles, USA\\
$^2$Department of Data Science and Artificial Intelligence, Monash University, Melbourne, Australia\\
$^3$Department of ECE, University of Wisconsin–Madison, Madison, USA\\
\emails
shaohung@usc.edu,
zhe.chen@monash.edu,
dianlun.lin@wisc.edu,
\{yue.zhang, daniel.harabor\}@monash.edu,
tsung-wei.huang@wisc.edu,
\{skoenig, thomy.phan\}@usc.edu
}

\begin{document}

\maketitle

\begin{abstract}
Multi-Agent Path Finding (MAPF) is the problem of finding a set of collision-free paths for multiple agents in a shared environment while minimizing the sum of travel time. Since solving the MAPF problem optimally is NP-hard, anytime algorithms based on Large Neighborhood Search (LNS) are promising to find good-quality solutions in a scalable way by iteratively destroying and repairing the paths. We propose \textit{Destroy-Repair Operation Parallelism for LNS (DROP-LNS)}, a parallel framework that performs multiple destroy and repair operations concurrently to explore more regions of the search space within a limited time budget. Unlike classic MAPF approaches, DROP-LNS can exploit parallelized hardware to improve the solution quality. We also formulate two variants of parallelism and conduct experimental evaluations. The results show that DROP-LNS significantly outperforms the state-of-the-art and the variants.
\end{abstract}

\section{Introduction}
A wide range of real-world applications can be formulated as \textit{Multi-Agent Path Finding (MAPF)} problem such as autonomous warehouse~\cite{WurmanAIM2008}, unmanned aerial vehicles~\cite{Ho2019MAPFUAV}, and autonomous vehicles~\cite{LiAAAI23Intersection}. MAPF aims to find a set of collision-free paths, each from an assigned start location to a goal location, for multiple agents in a shared environment while minimizing the sum of travel time~\cite{SternSoCS19MAPFDef}. However, solving MAPF optimally is NP-hard, which limits the scalability of many algorithms~\cite{YuAAAI2013NPHard}.

\textit{Anytime algorithms} are promising approaches to scale up MAPF to hundreds of agents by iteratively optimizing a set of collision-free paths until a user-specified time budget runs out. Based on \textit{Large Neighborhood Search (LNS)}, \textit{MAPF-LNS} is the current state-of-the-art algorithm in anytime MAPF~\cite{LiIJCAI21LNS}.
MAPF-LNS starts with an initial collision-free solution computed by a fast but suboptimal algorithm.
It iteratively selects a subset of agents, known as \textit{neighborhood}, and performs destroy and repair operations to replan their paths while keeping the other paths fixed. The repair operations can be done with any fast algorithm such as Prioritized Planning (PP)~\cite{SilverAAAI05PP}.
However, as the number of agents and the size of the environment increase, PP becomes a bottleneck during the search because the underlying single-agent path-finding algorithm slows down due to more temporal obstacles and longer path lengths. This predominantly affects the speed of the replan operations. Thus, MAPF-LNS may converge to poor-quality solutions in large-scale instances.

\begin{figure}[t!]
    \centering
    \begin{subfigure}[b]{0.32\linewidth}
        \centering
        \includegraphics[width=0.99\linewidth]{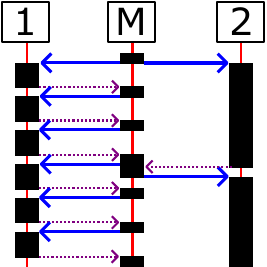}
        \caption{DROP-LNS}
        \label{fig:drop_lns_concept}
    \end{subfigure}
    \hfill
    \begin{subfigure}[b]{0.32\linewidth}
        \centering
        \includegraphics[width=0.99\linewidth]{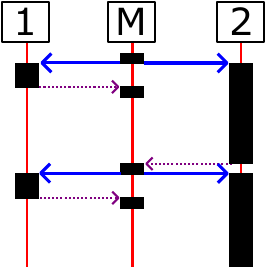}
        \caption{SYNC-LNS}
        \label{fig:sync_lns_concept}
    \end{subfigure}
    \hfill
    \begin{subfigure}[b]{0.32\linewidth}
        \centering
        \includegraphics[width=0.99\linewidth]{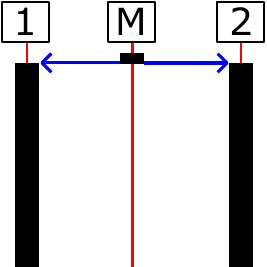}
        \caption{DETA-LNS}
        \label{fig:detached_lns_concept}
    \end{subfigure}
    \caption{Conceptual timelines of different parallelism variants with a main thread ``M'' and two worker threads ``1'' and ``2''. Black blocks indicate the \textit{productivity}, i.e., the time when threads are processing, and red lines indicate the idle time. Blue arrows indicate events where a worker thread receives tasks, and purple dotted arrows indicate events where a worker thread returns a solution.}
    \label{fig:concept}
\end{figure}%

Despite the rapid progress in the field of MAPF that resulted in many sophisticated algorithms~\cite{BoyarskiIJCAI15ICBS,LiAAAI2019RectangleReasoning}, few exploit parallelism to improve the anytime MAPF for better scalability and solution quality~\cite{LaurentNeurIPS2020Flatland}. Given the wide availability of parallel low-cost hardware, there is a lot of potential in leveraging parallelism to fully exploit the available time budget.
However, parallelization is not trivial as the search algorithm needs to balance between \textit{productivity} and \textit{synchronization}. The former describes the concurrent execution of tasks, and the latter describes the access to the best-known solution to guide the search toward better quality. Any imbalanced approach can lead to inefficient parallel search, where worker threads are either less productive due to \textit{synchronization overhead}, i.e., the idle time of the threads (see Figure~\ref{fig:sync_lns_concept}) or redundant because of exploring similar regions in the search space (see Figure~\ref{fig:detached_lns_concept}).

In this paper, we propose \textit{Destroy-Repair Operation Parallelism for LNS (DROP-LNS)}, a parallel framework to concurrently perform destroy and repair operations to explore more regions of the search space within a limited time budget. Unlike MAPF-LNS, DROP-LNS can exploit parallelized hardware to improve the solution quality.
Our contributions are as follows:%
\begin{itemize}
    \item We propose DROP-LNS, which performs pairs of destroy and repair operations concurrently on a solution via multi-threading. The solution is updated asynchronously for high productivity, i.e., less synchronization overhead.
    \item We formulate two other parallelism variants: SYNC-LNS, which synchronizes solutions at each iteration, and DETA-LNS, which detaches one MAPF-LNS to each thread and never synchronizes their solutions. We demonstrate that DROP-LNS offers a trade-off between productivity and synchronization in contrast to SYNC-LNS and DETA-LNS, as shown in Figure~\ref{fig:concept}.
    \item We evaluate DROP-LNS in six maps from the MAPF benchmark suite and demonstrate significantly better solution quality and scalability than the parallelism variants and the state-of-the-art.
\end{itemize}%

\section{Preliminaries}
\subsection{Problem Definition}
A MAPF problem consists of an undirected and unweighted graph $G = (V,E)$ and a set of $k$ \textit{agents} $A = \{a_1...a_k\}$, where $V$ is the set of vertices representing locations and $E$ is the set of edges. Each agent $a_i \in A$ has a start vertex $s_i$ and a goal vertex $g_i$. Time is discretized into timesteps.

A \textit{state} of an agent is represented as a tuple $(v,t)$ indicating its location at vertex $v$ at timestep $t$. At each timestep, an agent can either \textit{wait} at its current location or \textit{move} to an adjacent vertex. A \textit{collision} between two paths occurs when two agents occupy the same vertex or pass through the same edge in opposite directions at the same timestep.
A \textit{solution} is a set of paths $P = \{p_1,...,p_k\}$, with one path $p_i \in P$ for each agent $a_i \in A$.
A solution is \textit{feasible} if the set of paths is collision-free.
The \textit{cost} $c(p_i)$ of a solution $P$ for agent $a_i$ is the number of timesteps or travel time to get from start vertex $s_i$ to goal vertex $g_i$.  
In this paper, our goal is to find a feasible solution while minimizing the \textit{sum of costs (SOC)} $c(P) = \sum_{i=1}^{k}{c(p_i)}$.

\subsection{Large Neighborhood Search for MAPF}
To find solutions with low SOC within a user-specified time budget, anytime MAPF starts with a feasible solution that is iteratively optimized. The solution quality improves monotonically with increasing time budget ~\cite{cohen2018anytime,LiIJCAI21LNS}. Based on the \textit{Large Neighborhood Search (LNS)}, a meta-heuristic search algorithm for combinatorial optimization, \textit{MAPF-LNS} is the current state-of-the-art algorithm for anytime MAPF, which can scale up to large-scale scenarios with hundreds of agents \cite{HuangAAAI22MLLNS,LiIJCAI21LNS}. Starting with an initial feasible solution $P$ found by fast but suboptimal algorithms like PP~\cite{SilverAAAI05PP} or LaCAM~\cite{Okumura2023lacam2}, MAPF-LNS iteratively modifies $P$ by heuristically selecting a subset of $N$ agents $A' \subset A$ as the \textit{neighborhood} with their corresponding paths $P' = \{p_i \in P | a_i \in A'\}$ from $P$, where $N < k$ is a user-specified parameter. 
MAPF-LNS then repairs the paths $P'$ with a new set of paths $P'_\textit{new}$ generated by PP within a limited repair time budget. Since PP finds the paths sequentially according to a priority ordering of $A'$, each new path is supposed to avoid any collisions with the set of \textit{already-planned paths}, i.e., $(P \setminus P') \cup P'_\textit{new}$, to ensure feasibility of the solution. If MAPF-LNS finds such a path successfully, it adds the path to paths $P'_\textit{new}$. If each agent in neighborhood $A'$ has a corresponding path in paths $P'_\textit{new}$ and the SOC of the paths $P'_\textit{new}$ is lower than that of paths $P'$, then MAPF-LNS ``destroys'' the previous paths $P'$ from $P$ and ``repairs'' them with $P'_\textit{new}$.
We call the solution with the lowest SOC found so far during the search the \textit{best-known solution}. MAPF-LNS continues searching for solutions with lower SOCs until the time budget runs out and returns the latest best-known solution.

MAPF-LNS uses a set $\mathcal{H}$ of three \textit{destroy heuristics} for selecting neighborhoods, namely a \textit{random-based heuristic}, an \textit{agent-based heuristic}, and a \textit{map-based heuristic} \cite{LiIJCAI21LNS}.
% The \textit{random-based heuristic} uniformly selects $N$ random agents.
% The \textit{agent-based heuristic} generates the neighborhood based on the agent $a_i$ whose path has the maximum \textit{delay}, i.e., the difference between the timestep reaching its goal vertex and the shortest path length, and other agents that prevent $a_i$ from achieving a lower delay, which can be determined via random walks.
% The \textit{map-based heuristic} randomly chooses an intersection vertex $v \in V$ with a degree greater than two and generates a neighborhood of agents that traverse vertex $v$.
To select a destroy heuristic $H \in \mathcal{H}$ at each iteration, MAPF-LNS maintains a set of updatable \textit{weights} $w_H$, one for each destroy heuristic and uses a \textit{roulette wheel selection} mechanism, where each destroy heuristic $H$ is selected with the probability of $\frac{w_{H}}{\sum_{H \in \mathcal{H}}{w_H}}$~\cite{ropke2006adaptive,LiIJCAI21LNS}. After destroying and repairing paths of agents in the neighborhood, MAPF-LNS updates the value of weight $w_H$ corresponding to the selected destroy heuristic $H$:
\begin{equation}
    w_H \leftarrow \gamma \max\{c(P') - c(P'_\textit{new}), 0\} + (1-\gamma) w_H,
\end{equation}%
where $\gamma \in [0,1]$ is a user-specified reaction factor indicating how fast the weight value changes in reaction to the improvement of the solution quality.

\subsection{Lazy Constraint Addition Search for MAPF}
\textit{Lazy Constraint Addition Search for MAPF (LaCAM)}~\cite{Okumura2023lacam2} is a suboptimal algorithm that performs a search on the joint-state space, where each node in its search tree is a \textit{joint-state} of the agents on the graph, i.e., a sequence of non-repeated vertices, one for each agent.
To efficiently generate a suboptimal solution, LaCAM uses Priority Inheritance with Backtracking (PIBT)~\cite{OkumuraIJCAI2019PIBT}, a rule-based algorithm for solving MAPF suboptimally to determine movements for all agents step-wise.
When expanding a node, its successors are generated by invoking PIBT, with the move selected for each agent being restricted by a set of associated constraints (e.g., agents should not traverse the same vertex).
LaCAM systematically explores the set of joint-space nodes but uses partial expansion to mitigate the branching factor explosion. 
Adding a systematic search and the invocation of a pattern-based swap operation~\cite{luna2011push,de2014push} allows LaCAM to succeed more often than PIBT and compute higher-quality plans while retaining its performance advantages.

Based on LaCAM, \textit{LaCAM*}~\cite{Okumura2023lacam2} is an anytime algorithm that minimizes either the makespan (the maximum individual cost) or the sum of loss (the number of agents that have not reached their goals yet). Once LaCAM finds a solution, LaCAM* continues the search for better-quality solutions while keeping track of the minimum makespan (or sum of loss). We use LaCAM* as one of our baselines.

%%%%%%%%%%%%%%%%%%%%%%%%%%%%%%%%%%%%%%%%%%%%%%%%%%%%%%%%%%%%%%%%%%%%%%%%

\section{Related Work}
\subsection{Parallelism in MAPF}
Despite significant progress in MAPF in recent years, there has been limited work on parallelization to scale MAPF algorithms with the growing availability of low-cost parallel hardware.
Most works focus on some form of task decomposition, where parallelization is done at the level of the subproblems. Lee et al. decompose the task into subproblems solved in parallel and merge them into larger subproblems until the original task is completely solved~\cite{lee2021parallel}. Rahmani et al. decompose the map to perform a parallelized hierarchical pathfinding algorithm for multiple agents; however, they defer collision avoidance to the execution, which is different from our problem formulation~\cite{rahmani2020multi}. Caggianese et al. performed multiple single-agent path-finding processes in parallel, which also ignores collisions in between~\cite{CaggiansesICCS2012}.

In contrast to these works that focus on parallelization in finding one-shot solutions, we focus on iterative optimization in an anytime manner. Furthermore, we propose parallelization on the \textit{operation level} of the MAPF algorithm, which is less dependent on particular map structures and the number of agents.
The closest MAPF work to our focus is a naive variant, where multiple MAPF-LNS processes are performed on independent threads~\cite{LaurentNeurIPS2020Flatland}. When the time budget runs out, the best solution among all threads is used. However, this approach lacks a synchronization mechanism to focus the search on more promising solutions while avoiding wasteful overlaps in the independent search beams.%

\subsection{Parallel Adaptive Large Neighborhood Search}
Ropke extended LNS to \textit{Parallel Adaptive Large Neighborhood Search (PALNS)} to solve the Traveling Salesman Problem with Pickup and Delivery and the Capacitated Vehicle Routing Problem \cite{ropke2009parallel}. The tasks for parallelization comprise pairs of destroy and repair operators executed simultaneously on a given solution. Simulated annealing is used to decide whether a solution should be accepted as the solution for the next iteration or not.
% Shared variables include the solution to be processed, the solution for the next iteration, the destroy heuristic weights, the temperature parameter for simulated annealing, and an iteration counter.

We adopt a \textit{simplified variant} of PALNS for the MAPF setting, where we use the destroy heuristics proposed in \cite{LiIJCAI21LNS} and PP with randomized priorities as repair operator. Our approach does not use simulated annealing and is applied in an \textit{anytime manner} based on a time budget instead of an iteration count. Thus, it does not require the temperature and count value as additional shared variables.

\begin{figure}[t!]
    \centering
    \includegraphics[width=0.99\linewidth]{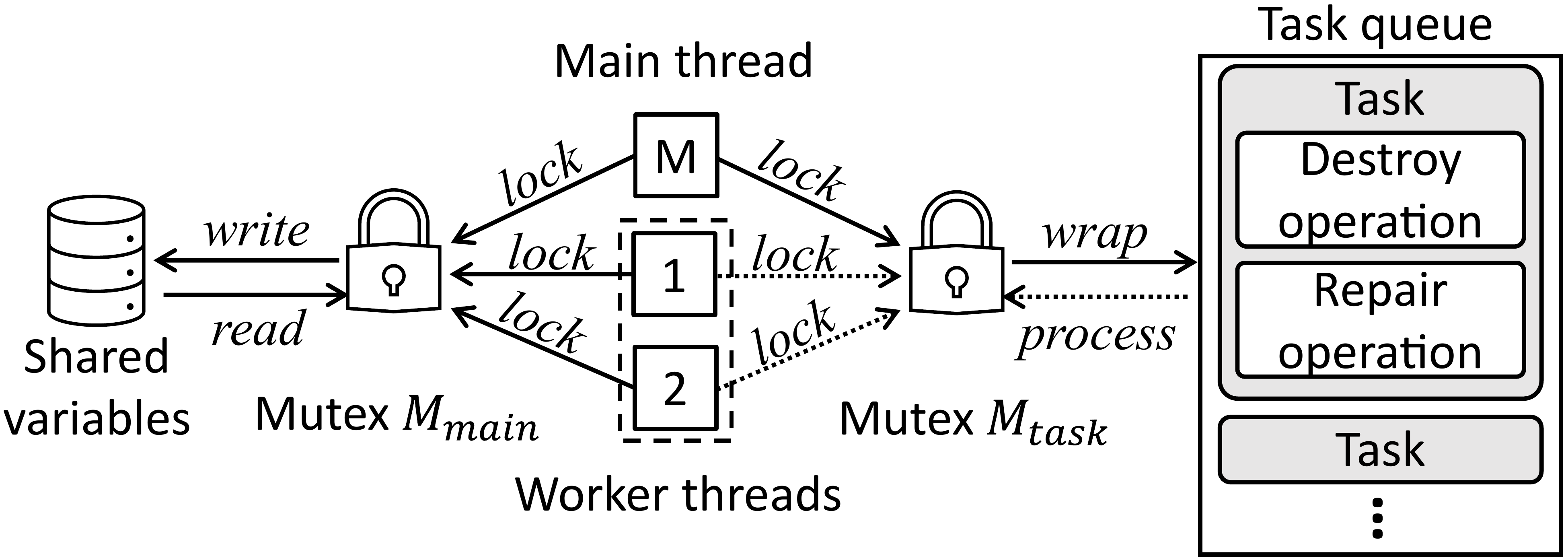}
    \caption{Illustrative example of the DROP-LNS framework with a main thread ``M" and two worker threads ``1" and ``2". Arrows are the actions from each thread.}
    \label{fig:system}
\end{figure}%

\section{Parallelism for MAPF-LNS}
We now introduce \textit{Destroy-Repair Operation Parallelism for MAPF-LNS (DROP-LNS)} and alternative approaches to parallelize MAPF-LNS.

\begin{algorithm}[tb!]
\caption{Main Function of DROP-LNS}\label{alg:MAIN}
\begin{algorithmic}[1]
\Procedure{Main}{$\,I, T, N, m\,$} %, dist, LB$}
\State Initialize \par
\State\hskip\algorithmicindent \textcolor{blue}{$\mathbf{w}_\textit{H} \leftarrow 1, \forall H \in \mathcal{H}$} \par 
\State\hskip\algorithmicindent \textcolor{blue}{task queue $\mathbf{Q}_\textit{task} \leftarrow \phi$} \par
\State\hskip\algorithmicindent \textcolor{blue}{mutexes $M_\textit{main}$ and $M_\textit{task}$} \par
\State \textcolor{blue}{$\mathbf{P}_\textit{min} = \{ p_1, ..., p_k \} \leftarrow$ Find initial solution with \textit{I}, \textit{T}} %Run MAPF algorithm with \textit{I} and \textit{T} 
\State Activate $m$ worker threads
\While{runtime \textbf{not} exceeds $T$}
    \State \textcolor{blue}{[lock $M_\textit{task}$]}
    \If{\textcolor{blue}{$\mathbf{Q}_\textit{task}$ \textbf{not} full}}
        \State \textcolor{blue}{Wrap \textsc{DestroyAndRepair} tasks to $\mathbf{Q}_\textit{task}$}
    \EndIf
    \State \textcolor{blue}{[unlock $M_\textit{task}$]}
\EndWhile
\State Stop all the worker threads
\State \Return $\mathbf{P}_\textit{min}$
\EndProcedure
\end{algorithmic}
\end{algorithm}%

\subsection{Destroy-Repair Operation Parallelism}
Figure~\ref{fig:system} shows an illustrative example of the DROP-LNS framework.
DROP-LNS uses a \textit{main thread} and $m$ \textit{worker threads} to parallelize the search. The core idea is to wrap pairs of destroy and repair operations as \textit{tasks} via the main thread and assign these tasks to idle worker threads. DROP-LNS maintains \textit{shared variables}, denoted in bold, that can be modified by any thread, including the \textit{best-known solution} $\mathbf{P}_\textit{min}$ with the minimum SOC found so far, the weights $\mathbf{w}_\textit{H}$ for each destroy heuristic $H \in \mathcal{H}$, and a task queue $\mathbf{Q}_\textit{task}$ with a fixed capacity. It also maintains \textit{shared constants} that are read-only and cannot be modified after initialization, including the given MAPF instance $I$ and three user-specified parameters: the neighborhood size $N$, the time budget $T$, and the reaction factor $\gamma$. To synchronize the shared variables, DROP-LNS uses two \textit{mutexes}: $M_{\textit{task}}$ to ensure that only one thread is allowed to modify the task queue at a time, and $M_{\textit{main}}$ for modifying any other shared variables.
Any thread should acquire or \textit{lock} the mutex to modify the corresponding shared variables and release that mutex after the modification. For example, once a worker thread finishes its current task, it tries to acquire mutex $M_{\textit{task}}$ to pop the next task out of task queue $\mathbf{Q}_\textit{task}$. If mutex $M_{\textit{task}}$ is locked by another thread, the worker thread has to wait for it to be released. Analogously, mutex $M_{\textit{main}}$ has to be acquired to access or modify the other shared variables.

Like MAPF-LNS, DROP-LNS first uses the main thread to initialize a feasible solution via LaCAM~\cite{Okumura2023lacam2}. The main thread then fills task queue $\mathbf{Q}_\textit{task}$ with the fixed capacity of tasks until the time budget runs out. The process is formulated in Algorithm~\ref{alg:MAIN}, where $I$ is the MAPF instance to be solved, $T$ is the time budget, $N$ is the neighborhood size, and $m$ is the number of threads.
We mark lines requiring access to shared variables in blue.

An idle worker thread tries to access task queue $\mathbf{Q}_\textit{task}$ by acquiring mutex $M_\textit{task}$. If successful, the thread locks $M_\textit{task}$ and pops a task from the queue. The task function of DROP-LNS is formulated in Algorithm~\ref{alg:TASK}, where each worker thread performs a task with private variables that can only be accessed and modified by the corresponding worker thread.
Before executing a task, the worker thread tries to acquire mutex $M_\textit{main}$ in order to copy the shared variables to its private variables, including a copy $P$ of the currently best-known solution $\mathbf{P}_\textit{min}$ and a copy $w_\textit{H}$ of weights $\mathbf{w_\textit{H}}$, for all $H \in \mathcal{H}$. The worker thread then releases mutex $M_\textit{main}$ to enable access or modification of the shared variables by other threads and computes the SOC $C$ of paths $P$ [Lines~\ref{alg2:initial_start}-\ref{alg2:initial_end}]. 
To perform destroy operations, the worker thread samples a destroy heuristic $H'$ using the roulette wheel selection mechanism with weights $w_\textit{H}$. Then, the worker thread selects a subset of $N$ agents $A' \subseteq A$ as the neighborhood along with their paths $P' \subseteq P$ according to destroy heuristic $H'$, and removes $P'$ from $P$ [Lines~\ref{alg2:destroy_start}-\ref{alg2:destroy_end}].

\begin{algorithm}[t!]
\caption{Task Function of DROP-LNS}\label{alg:TASK}
\begin{algorithmic}[1]
\Procedure{DestroyAndRepair}{}
\State Initialize \par \label{alg2:initial_start}
\State\hskip\algorithmicindent \textcolor{blue}{[lock $M_\textit{main}$]} \par
\State\hskip\algorithmicindent \textcolor{blue}{$P \:\; \leftarrow \mathbf{P}_\textit{min}$} \par
\State\hskip\algorithmicindent \textcolor{blue}{$w_\textit{H} \leftarrow \mathbf{w}_\textit{H}, \; \forall H \in \mathcal{H}$} \par
\State\hskip\algorithmicindent \textcolor{blue}{[unlock $M_\textit{main}$]} \par
\State\hskip\algorithmicindent $C \; \; \leftarrow c(P)$ \label{alg2:initial_end}
\State $H' \;\; \leftarrow$ Select a destroy heuristic with $w_\textit{H},\, \forall H \in \mathcal{H}$ \label{alg2:destroy_start}
\State $A' \,\;\; \leftarrow$ Select a neighborhood with $H'$
\State $P' \,\;\; \leftarrow \{p_i \in P | a_i \in A'\}$
\State $P \,\;\;\; \leftarrow P \setminus P'$ \Comment{Destroy the subset of paths}  \label{alg2:destroy_end}
\State $P'_\textit{new} \leftarrow$ Run PP within a repair time budget \label{alg2:fail_start}
\If{$P'_\textit{new}$ \textbf{not} found \textbf{or} $c(P'_\textit{new}) \geq c(P')$}
    \State \textcolor{blue}{[lock $M_\textit{main}$]}
    \State \textcolor{blue}{$\mathbf{w}_{\textit{H}_{\textit{s}}} \leftarrow (1-\gamma) \cdot w_{\textit{H}_{\textit{s}}}$}
    \State \textcolor{blue}{[unlock $M_\textit{main}$]}
    \State \Return
\EndIf \label{alg2:fail_end}
\If{runtime \textbf{not} exceeds $T$} \label{alg2:succ_start}
    \LComment{The worker thread finds better paths within $T$}
    \State $P \leftarrow P \cup P'_\textit{new}$ \Comment{Repair the subset of paths}
    \State \textcolor{blue}{[lock $M_\textit{main}$]}
    \State \textcolor{blue}{$\mathbf{w}_{\textit{H}_{\textit{s}}} \leftarrow \gamma \cdot (C - c(P)) + (1-\gamma) \cdot w_{\textit{H}_{\textit{s}}}$}
    \If{\textcolor{blue}{$c(P) < c(\mathbf{P}_\textit{min})$}}
        \State \textcolor{blue}{$\mathbf{P}_\textit{min} \leftarrow P$}
    \EndIf
    \State \textcolor{blue}{[unlock $M_\textit{main}$]}
\EndIf  \label{alg2:succ_end}
\State \Return
\EndProcedure
\end{algorithmic}
\end{algorithm}%

To perform the repair operation, the worker thread uses PP to generate paths for agents in neighborhood $N$. If PP fails to find such paths within the repair time budget or the repaired paths $P'_\textit{new}$ has a higher SOC than that of destroyed paths $P'$, the worker thread acquires mutex $M_\textit{main}$ and lowers weight $\mathbf{w}_{H'}$ by a factor of $1-\gamma$ without updating paths $\mathbf{P}_\textit{min}$, which indicates a failed iteration. After that, the worker thread terminates the task function and tries to fetch the first task in task queue $\mathbf{Q}_\textit{task}$ again [Lines~\ref{alg2:fail_start}-\ref{alg2:fail_end}].
Otherwise, the worker thread repairs paths $P$ with $P'_\textit{new}$ and acquires mutex $M_\textit{main}$ in order to access the shared variables, namely paths $\mathbf{P}_\textit{min}$ and weight $\mathbf{w}$. The worker thread first updates the value of weight $\mathbf{w}_{H'}$ to
\begin{equation}
    \mathbf{w}_{H'} \leftarrow \gamma [c(P) - c(P\setminus P' \cup P'_\textit{new})] + (1-\gamma) w_{H'},
\end{equation}%
as it modifies the SOC from $c(P)$ to $c(P\setminus P' \cup P'_\textit{new})$ after processing the task. Then, the worker thread updates path $\mathbf{P}_\textit{min}$ if SOC $c(P\setminus P' \cup P'_\textit{new}))$ is lower than that of $c(P_\textit{min})$, and releases the mutex so that other threads can access the shared variables [Lines~\ref{alg2:succ_start}-\ref{alg2:succ_end}]. We define \textit{synchronization} as a two-step process that (1) compares the SOCs between the solution $P$ generated by the worker thread and the best-known solution $\mathbf{P}_\textit{min}$ and (2) updates the best-known solution and the weight of each destroy heuristic accordingly.

\subsection{Parallelism Variants of MAPF-LNS}
We also implement two other parallelism variants: \textit{SYNC-LNS} and \textit{DETA-LNS}.

\paragraph{\textbf{SYNC-LNS}}
SYNC-LNS keeps track of the best-known solution $\mathbf{P}_\textit{min}$. At each iteration, SYNC-LNS selects the same number of neighborhoods as the worker threads. Each worker thread then performs a pair of destroy and repair operations individually in parallel. After all the worker threads complete their own destroy and repair operations, SYNC-LNS selects the worker thread that contains the solution $P_\textit{min}$ with the lowest SOC among all solutions generated by all worker threads. It also records the destroy heuristic $H' \in \mathcal{H}$ that the selected worker thread uses.
SYNC-LNS compares the SOC between the solution $P_\textit{min}$ and the best-known solution $\mathbf{P}_\textit{min}$, and then updates the value of weight $w_{H'}$ to
\begin{equation}
    w_{H'} \leftarrow \gamma \max\{c(\mathbf{P}_\textit{min}) - c(P_\textit{min}), 0\} + (1-\gamma) w_{H'}.
\end{equation}%
SYNC-LNS replaces solution $\mathbf{P}_\textit{min}$ with $P_\textit{min}$ if the latter has a lower SOC than the former.
That is, it synchronizes solutions and weights at each iteration by selecting the one with the lowest SOC, as illustrated in Figure \ref{fig:sync_lns_concept}.

\paragraph{\textbf{DETA-LNS}}
Inspired by~\cite{LaurentNeurIPS2020Flatland} that processes LNS using four CPUs, DETA-LNS ``detaches'' one MAPF-LNS process individually on a worker thread in parallel with equal weights of each destroy heuristic. When the time budget runs out, it selects the solution with the lowest SOC among all solutions generated by the worker threads. That is, DETA-LNS never synchronizes solutions and weights developed by each thread until the time budget runs out, as illustrated in Figure \ref{fig:detached_lns_concept}.

\subsection{Conceptual Discussion}
DROP-LNS parallelizes the search by wrapping pairs of destroy and repair operations as concurrently executable tasks while maintaining the best-known solution and the weights for destroy heuristic selection.
Compared to performing destroy and repair operations sequentially on a single worker thread, parallelism can efficiently exploit high-quality solutions for further improvement and explore different regions in the search space. Both aspects increase the chance of finding solutions with lower SOCs than MAPF-LNS.

Figure~\ref{fig:concept} shows the conceptual timelines during the search of DROP-LNS, SYNC-LNS, and DETA-LNS, respectively.
Regarding synchronization, SYNC-LNS performs a focused search as it prunes solutions with higher SOCs at each iteration, meaning that all worker threads can focus on higher-quality solutions. However, SYNC-LNS must wait until all worker threads complete their tasks before moving on to the next iteration, which limits the productivity of fast worker threads and thus increases the synchronization overhead.

On the other hand, DETA-LNS only synchronizes solutions at the end when the time budget runs out, meaning that its worker threads do not need to wait for one another. Thus, DETA-LNS theoretically exhibits the highest possible productivity of worker threads. Since all worker threads process independently, they may explore overlapping regions in the search space, which leads to finding similar solutions and thus limits the effectiveness due to a lack of exploitation of high-quality solutions.

As shown in Figure \ref{fig:drop_lns_concept}, DROP-LNS synchronizes solutions on the fly. Once a worker thread completes its task, it synchronizes without waiting for others. Thus, DROP-LNS maintains higher productivity than SYNC-LNS. Unlike DETA-LNS, DROP-LNS still synchronizes its solutions (albeit at the risk of some worker threads wasting some computation on outdated solutions).
Thus, DROP-LNS trades off between pruning bad-quality solutions and the synchronization overhead.

\section{Empirical Evaluation}
\subsection{Configurations and Algorithm Implementation}
\begin{figure*}[t!]
    \centering
    \includegraphics[width=0.99\linewidth]{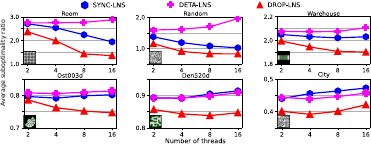}
    \caption{Average solution quality among all instances with the highest number of agents on each map solved by SYNC-LNS, DETA-LNS, and DROP-LNS with 2, 4, 8, and 16 threads, respectively. The lower the average suboptimality ratio, the better the solution quality.}
    \label{fig:threads}
\end{figure*}

\begin{figure*}[t!]
    \centering
    \includegraphics[width=0.99\linewidth]{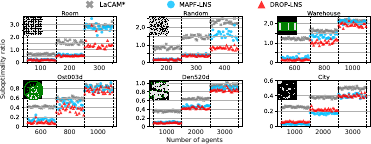}
    \caption{Solution quality of instances solved by LaCAM*, MAPF-LNS, and DROP-LNS. Instances are grouped by the number of agents. The lower the average suboptimality ratio, the better the solution quality.}
    \label{fig:sota_performance}
\end{figure*}

% \paragraph{\textbf{Configurations of MAPF Instances}}
We evaluate DROP-LNS on six 4-connected grid maps from the MAPF benchmark suite~\cite{SternSoCS19MAPFDef}, namely (1) a \texttt{Room} map (\textit{room-32-32-4}) of size $32 \times 32$, (2) a \texttt{Random} map with $20\%$ static obstacles (\textit{random-32-32-10}) of size $32 \times 32$, (3) a \texttt{Warehouse} map (\textit{warehouse-10-20-10-2-1}) of size $161 \times 63$, two maps from the game Dragon Age: Origins, which are (4) \texttt{Ost003d} of size $194 \times 194$ and (5) \texttt{Den520d} of size $256 \times 257$, as well as (6) a \texttt{City} map (\textit{Paris\_1\_256}) of size $256 \times 256$. Figures~\ref{fig:threads} and~\ref{fig:sota_performance} show the configuration of each map. We conduct all experiments on the available 25 random scenarios for each map. Since the benchmark suite provides only instances with at most 1,000 agents, we create 25 instances, each with 2000 and 3000 agents for \texttt{Den520d} and \texttt{City} maps. The start and goal vertices of all agents are randomly selected. All the experiments are conducted within a time budget $T=60$ seconds.

% \paragraph{\textbf{Algorithm Implementation}}
We implement DROP-LNS, SYNC-LNS, and DETA-LNS as parallelism variants with fixed neighborhood size $\textit{N} = 16$ and reaction factor $\gamma = 0.01$ and use $m=8$ worker threads unless mentioned otherwise.
We adopt the public implementations of MAPF-LNS (with the same $N$ and $\gamma$)~\cite{LiIJCAI21LNS} and LaCAM*~\cite{Okumura2023lacam2}.
We modify LaCAM* by tracking the total number of timesteps agents took before their last visit to their respective goal vertices so that it becomes an anytime algorithm that optimizes SOC.
We implement all algorithms in C++ (compiled with GCC-11.3.0) and run experiments on CentOS Linux and an AMD EPYC 7302 16-core processor with 16 GBs of memory.

\subsection{Evaluation Metrics}
\paragraph{\textbf{Solution Quality}}
To evaluate the solution quality among instances with various numbers of agents and sizes of graphs, we first define \textit{the shortest path distance} $d_i$ of an agent $a_i$ as the minimum timesteps needed to move from start vertex $s_i$ to goal vertex $g_i$. 
Thus, the sum of the shortest path distances provides a \textit{lowerbound} of the optimal SOC. We define the \textit{delay} of an agent $a_i$ as the timestep difference between its path $p_i$ and its shortest path distance $d_i$, i.e., $c(p_i) - d_i$. To evaluate the solution quality of different algorithms, we compare the \textit{suboptimality ratio} between the \textit{sum of delays} over all agents and the lowerbound
\begin{equation}
    \textit{suboptimality ratio} = \frac{\sum_{a_i \in A}{(c(p_i) - d_i)}}{\sum_{a_i \in A}{d_i}}.
\end{equation}%
Since the sum of the shortest path distances is a constant given an instance, comparing the suboptimality ratios is equivalent to comparing the SOCs of solutions. The lower the suboptimality ratio, the better the solution quality is.

\begin{figure}[tb!]
    \centering
    \includegraphics[width=0.99\linewidth]{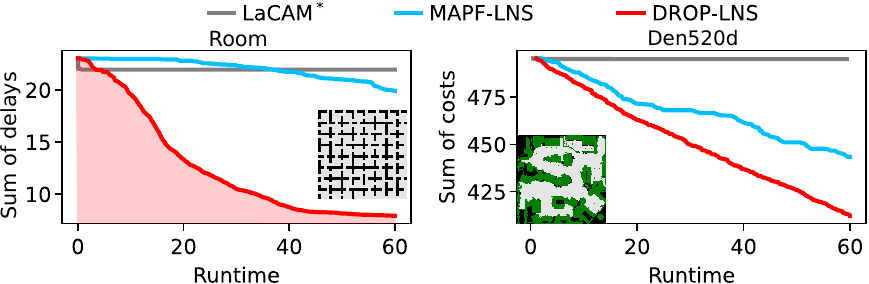}
    \caption{Sum of delays (in thousands) versus runtime (in seconds) in two instances with 300 agents in \texttt{Room} map and with 3000 agents in \texttt{Den520d} respectively solved by LaCAM*, MAPF-LNS, and DROP-LNS. The pink region indicates the AUC of DROP-LNS.}
    \label{fig:auc_example}
\end{figure}

\paragraph{\textbf{Effectiveness}}
To evaluate the effectiveness of algorithms, we focus on the following aspects.
\begin{itemize}
    \item[\textbf{(1)}] How fast an algorithm can find a high-quality solution.
    \item[\textbf{(2)}] How productive the worker threads are.
    \item[\textbf{(3)}] How often an algorithm exploits the high-quality solutions found so far during the search.
    \item[\textbf{(4)}] How often an algorithm explores different solutions.
\end{itemize}%
For aspect \textbf{(1)}, we measure the \textit{area under curve} (AUC), which is defined as the integral of the sum of delays of the best-known solution versus runtime (see Figure~\ref{fig:auc_example}). The lower the AUC, the faster an algorithm converges to a high-quality solution. For aspect \textbf{(2)}, we measure the \textit{number of pairs of destroy and repair operations} (NPO) during the search, where NPO* denotes the total NPO within the time budget. The higher the NPO*, the more productive the worker threads are (i.e., less synchronization overhead). For aspect \textbf{(3)}, we measure the \textit{depth} of the final solution (DP), which is defined as the NPO from the initial solution to the final best-known solution. The higher the DP, the more frequently the algorithm improves its best-known solution. For aspect \textbf{(4)}, we measure the \textit{exploration ratio} (EXP), which is defined as the ratio between the NPO that does not lead to the final best-known solution and the total NPO, i.e.,
\begin{equation}
\text{EXP} = \frac{\text{NPO*}-\text{DP}}{\text{NPO*}}.
\end{equation}%
The higher the EXP, the more frequently the algorithm explores in the search space.

\begin{table*}[tb!]
\centering
\small
\tabcolsep=0.095cm
\renewcommand{\arraystretch}{1.2}
\begin{tabular}{|l|r|r|rrrr|rrrr|rrrr|rrrr|}
\hline
\multirow{3}{*}{}   & \multicolumn{1}{c|}{\multirow{3}{*}{$k$}} & \multicolumn{1}{c|}{\multirow{2}{*}{LaCAM*}} & \multicolumn{4}{c|}{\multirow{2}{*}{MAPF-LNS}}                                                         & \multicolumn{4}{c|}{\multirow{2}{*}{SYNC-LNS}}                                                         & \multicolumn{4}{c|}{\multirow{2}{*}{DETA-LNS}}                                                         & \multicolumn{4}{c|}{\multirow{2}{*}{DROP-LNS}}                                                         \\
                    & \multicolumn{1}{c|}{}                     & \multicolumn{1}{c|}{}                        & \multicolumn{4}{c|}{}                                                                                  & \multicolumn{4}{c|}{}                                                                                  & \multicolumn{4}{c|}{}                                                                                  & \multicolumn{4}{c|}{}                                                                                  \\ \cline{3-19} 
                    & \multicolumn{1}{c|}{}                     & \multicolumn{1}{r|}{AUC}                     & \multicolumn{1}{c}{AUC} & \multicolumn{1}{c}{NPO*} & \multicolumn{1}{c}{DP} & \multicolumn{1}{c|}{EXP} & \multicolumn{1}{c}{AUC} & \multicolumn{1}{c}{NPO*} & \multicolumn{1}{c}{DP} & \multicolumn{1}{c|}{EXP} & \multicolumn{1}{c}{AUC} & \multicolumn{1}{c}{NPO*} & \multicolumn{1}{c}{DP} & \multicolumn{1}{c|}{EXP} & \multicolumn{1}{c}{AUC} & \multicolumn{1}{c}{NPO*} & \multicolumn{1}{c}{DP} & \multicolumn{1}{c|}{EXP} \\ \hline
\multirow{3}{*}{\texttt{RO}} & 100                                       & 0.10                                         & \textbf{0.03}           & 26.6                     & 137.1                  & \textbf{0.99}                     & \textbf{0.03}           & 72.0                     & 127.5                  & \textbf{0.99}            & \textbf{0.03}           & \textbf{139.4}           & \textbf{142.5}         & \textbf{0.99}            & \textbf{0.03}           & 129.2                    & 134.9                  & \textbf{0.99}            \\
                    & 200                                       & 0.48                                         & 0.24                    & 16.9                     & 547.3                  & 0.96                     & 0.20                    & 62.7                     & 562.7                  & \textbf{0.99}            & 0.25                    & 86.6                     & 522.1                  & \textbf{0.99}            & \textbf{0.19}           & \textbf{104.6}           & \textbf{628.7}         & \textbf{0.99}            \\
                    & 300                                       & 1.32                                         & 1.32                    & 8.0                      & 68.9                   & \textbf{0.99}                     & 1.23                    & 27.4                     & 323.0                  & \textbf{0.99}                     & 1.34                    & 47.2                     & 52.0                   & \textbf{0.99}            & \textbf{1.00}           & \textbf{62.0}            & \textbf{1194.1}        & 0.98                     \\ \hline
\multirow{3}{*}{\texttt{RA}} & 200                                       & 0.22                                         & \textbf{0.05}           & 21.1                     & 372.1                  & 0.98                     & \textbf{0.05}           & 57.1                     & 308.7                  & \textbf{0.99}                     & 0.06                    & \textbf{116.4}                    & \textbf{386.3}         & \textbf{0.99}            & \textbf{0.05}           & 108.5           & 366.9                  & \textbf{0.99}            \\
                    & 300                                       & 0.57                                         & 0.23                    & 15.7                     & 885.2                  & 0.94                     & 0.21                    & 50.7                     & 763.2                  & 0.98                     & 0.26                    & 80.8                     & 835.0                  & \textbf{0.99}            & \textbf{0.19}           & \textbf{95.1}            & \textbf{976.6}         & \textbf{0.99}            \\
                    & 400                                       & 1.22                                         & 1.08                    & 6.5                      & 497.4                  & 0.92                     & 0.83                    & 31.4                     & 1158.8                 & 0.96                     & 1.12                    & 36.3                     & 375.9                  & \textbf{0.99}            & \textbf{0.69}           & \textbf{70.1}            & \textbf{2063.7}        & 0.97                     \\ \hline
\multirow{3}{*}{\texttt{W}}  & 600                                       & 3.48                                         & 2.02                    & 0.8                      & 560.7                  & 0.26                     & 2.52                    & 1.1                      & 144.76                 & \textbf{0.88}            & 2.69                    & 1.6                      & 221.3                  & 0.87                     & \textbf{1.51}           & \textbf{8.0}             & \textbf{968.6}         & 0.87                     \\
                    & 800                                       & 6.19                                         & 5.78                    & 0.1                      & 124.9                  & 0.13                     & 5.88                    & 0.4                      & 54.9                   & 0.87                     & 6.00                    & 0.6                      & 74.2                   & \textbf{0.88}            & \textbf{5.20}           & \textbf{1.2}             & \textbf{227.7}         & 0.80                     \\
                    & 1000                                      & 10.09                                        & 10.11                   & 0.1                      & 53.4                   & 0.31                     & 10.06                   & 0.3                      & 36.1                   & 0.88                     & 10.19                   & 0.4                      & 38.0                   & \textbf{0.90}            & \textbf{9.74}           & \textbf{0.6}             & \textbf{85.6}          & 0.85                     \\ \hline
\multirow{3}{*}{\texttt{O}}  & 600                                       & 2.24                                         & 1.23                    & 0.4                      & 285.0                  & 0.32                     & 1.33                    & 1.2                      & 146.7                  & 0.88                     & 1.50                    & 1.9                      & 198.9                  & \textbf{0.89}            & \textbf{0.93}           & \textbf{3.9}             & \textbf{459.2}         & 0.88                     \\
                    & 800                                       & 4.23                                         & 3.93                    & 0.1                      & 81.5                   & 0.28                     & 3.93                    & 0.3                      & 43.0                   & 0.88                     & 4.04                    & 0.5                      & 49.8                   & \textbf{0.90}            & \textbf{3.40}           & \textbf{0.8}             & \textbf{136.2}         & 0.82                     \\
                    & 1000                                      & 7.51                                         & 7.53                    & 0.1                      & 28.3                   & 0.45                     & 7.54                    & 0.2                      & 18.8                   & 0.88                     & 7.61                    & 0.2                      & 18.6                   & \textbf{0.92}            & \textbf{7.26}           & \textbf{0.3}             & \textbf{47.6}          & 0.83                     \\ \hline
\multirow{3}{*}{\texttt{D}}  & 1000                                      & 3.61                                         & 1.68                    & 0.7                      & \textbf{514.1}         & 0.25                     & 1.99                    & 1.7                      & 206.6                  & 0.88                     & 2.03                    & \textbf{3.5}                      & 372.5                  & \textbf{0.89}            & \textbf{1.66}           & 3.3             & 414.5                  & 0.87                     \\
                    & 2000                                      & 13.21                                        & 11.46                   & 0.2                      & \textbf{220.1}         & 0.03                     & 12.30                   & 0.5                      & 64.2                   & \textbf{0.88}            & 12.07                   & 1.0                      & 144.0                  & 0.86                     & \textbf{11.04}          & \textbf{1.3}             & 199.2                  & 0.85                     \\
                    & 3000                                      & 29.41                                        & 28.31                   & 0.1                      & 100.3                  & 0.06                     & 28.76                   & 0.3                      & 37.52                  & \textbf{0.88}            & 28.47                   & 0.6                      & 77.0                   & 0.87            & \textbf{27.73}          & \textbf{0.7}             & \textbf{114.0}         & 0.84                     \\ \hline
\multirow{3}{*}{\texttt{C}}  & 1000                                      & 2.83                                         & \textbf{0.77}           & 1.7                      & \textbf{766.4}         & 0.55                     & 1.18                    & 2.5                      & 294.9                  & 0.88                     & 1.22                    & \textbf{5.9}             & 487.8                  & \textbf{0.92}            & 1.24                    & 2.1                      & 344.4                  & 0.83                     \\
                    & 2000                                      & 8.47                                         & \textbf{5.78}           & 0.6                      & \textbf{572.6}         & 0.06                     & 7.49                    & 0.8                      & 104.1                  & \textbf{0.88}            & 6.82                    & \textbf{2.1}             & 296.3                  & 0.86                     & 6.57                    & 1.6                      & 248.2                  & 0.84                     \\
                    & 3000                                      & 16.58                                        & 15.10                   & 0.3                      & \textbf{258.2}         & 0.05                     & 16.25                   & 0.5                      & 58.8                   & \textbf{0.88}            & 15.65                   & \textbf{1.1}             & 156.8                  & 0.85                     & \textbf{15.08}          & \textbf{1.1}             & 193.8                  & 0.83                     \\ \hline
\end{tabular}%
\caption{Average AUC (in millions), NPO* (in thousands), DP, and EXP over all instances with the same number of agents per map. \texttt{RO}, \texttt{RA}, \texttt{W}, \texttt{O}, \texttt{D}, and \texttt{C} in the first column stand for maps \texttt{Room}, \texttt{Random}, \texttt{Warehouse}, \texttt{Ost003d}, \texttt{Den520d}, and \texttt{City}, respectively. For AUC (w.r.t. NPO*, DP, and EXP), numbers in bold are the minimum (w.r.t. maximum) among all in the same row.}
\label{tab:auc_all}
\end{table*}%

\begin{figure}[tb!]
    \centering
    \includegraphics[width=0.99\linewidth]{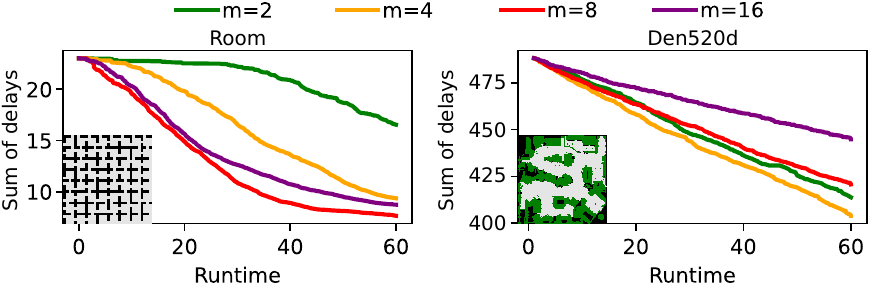}
    \caption{Sum of delays (in thousands) versus runtime (in second) in two instances with 300 agents in \texttt{Room} maps and with 3000 agents in \texttt{Den520d} respectively. Each instance is solved by DROP-LNS with $m=2,4,8,16$ worker threads.}
    \label{fig:threads_iter_drop}
\end{figure}%

\begin{figure}[tb!]
    \centering
    \includegraphics[width=0.99\linewidth]{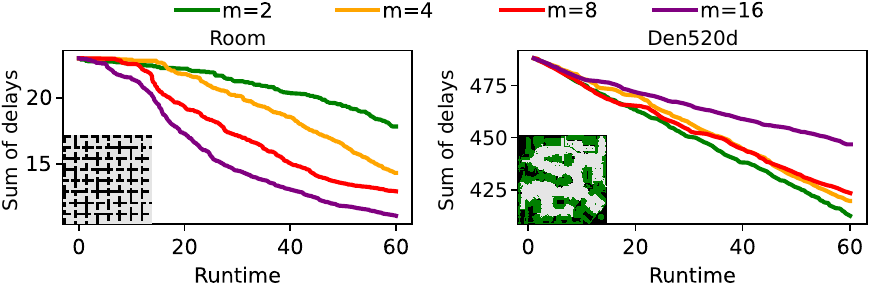}
    \caption{Sum of delays (in thousands) versus runtime (in second) in two same instances as in~\ref{fig:threads_iter_drop}. Each instance is solved by SYNC-LNS with $m=2,4,8,16$ worker threads.}
    \label{fig:threads_iter_sync}
\end{figure}%

\begin{table}[tb!]
\centering
\small
\tabcolsep=0.11cm
\renewcommand{\arraystretch}{1.4}
\begin{tabular}{|l|rrrrrr|}
\hline
            & \texttt{RO} & \texttt{RA} & \texttt{W} & \texttt{O} & \texttt{D} & \texttt{C} \\ \hline
LaCAM*      &      3723.4 &   4567.2    &    2491.4  &    3260.7  &  2664.0    &  2733.6    \\
MAPF-LNS    &         8.3 &     10.8    &     118.7  &     249.8  &   802.0    &   940.8    \\
SYNC-LNS    &         8.4 &      9.5    &      67.0  &     181.0  &   786.9    &   942.5    \\
DETA-LNS    &        13.0 &     15.7    &     137.1  &     315.8  &  1096.3    &  1449.3    \\
DROP-LNS    &         9.8 &     11.1    &     145.5  &     382.2  &  1294.1    &  1703.5    \\ \hline
\end{tabular}
\caption{Average memory usage of algorithms over instances with the highest number of agents on each map in MB. Labels in the first row indicate the same maps as Table~\ref{tab:auc_all}.}
\label{tab:rss}
\end{table}%

\subsection{Empirical Results}
Figure~\ref{fig:threads} shows the average solution quality of different parallelism variants versus the number of threads, where DROP-LNS outperforms SYNC-LNS and DETA-LNS while maintaining its performance as the number of threads increases.
Figure~\ref{fig:sota_performance} shows the solution quality of DROP-LNS along with the state-of-the-art algorithms for anytime MAPF, namely LaCAM* and MAPF-LNS. DROP-LNS outperforms the state-of-the-art algorithms, especially in small and congested instances such as \texttt{Room} or \texttt{Random} maps with increasing numbers of agents.
Figure~\ref{fig:auc_example} shows the changes in the sum of delays versus the runtime while solving an instance where DROP-LNS converges to a good-quality solution faster than the state-of-the-art algorithms, resulting in lower AUC.
Figure~\ref{fig:threads_iter_drop} shows an example of how DROP-LNS with different numbers of threads converges to good-quality solutions during the search. In congested instances such as \texttt{Room}, DROP-LNS converges to good-quality solutions faster as the number of threads increases from 2 to 8 but gets slower when increasing from 8 to 16 due to the synchronization overhead. The synchronization overhead becomes more significant when solving instances with large maps such as \texttt{Den520d}, where DROP-LNS converges faster only when the number of threads increases from 2 to 4 but gets slower afterward.
Figure~\ref{fig:threads_iter_sync} shows how SYNC-LNS converges, where its sum of delays is larger than DROP-LNS. In instances with large maps such as \texttt{Den520d}, SYNC-LNS converges slower as the number of threads increases from 2, showing its high synchronization overhead.
More overall statistic results are shown in Table~\ref{tab:auc_all}.
Table~\ref{tab:auc_all} shows the average AUC, NPO*, DP, and EXP among all compared algorithms. The AUC and DP of DETA-LNS are obtained from the worker thread that has the best-known solution when the search ends. DROP-LNS typically has lower AUC, especially in small and congested instances.
Table~\ref{tab:rss} shows the average memory usage over instances on the same map, where LaCAM* can be more memory-consuming than all the MAPF-LNS variants in two orders of magnitude when solving congested instances.

\subsection{Empirical Discussion}
Repair operations in large maps such as \texttt{Den520d} and \texttt{City} can be time-consuming due to the long distance between start and goal vertices. Thus, the NOP* is lower than in small maps, and parallelism becomes less effective. Also, the agents are less congested in large maps, meaning that one agent may have a near-optimal path without colliding with others. Thus, the initial solution provided by LaCAM is already near-optimal, limiting the effectiveness of parallelism.

DETA-LNS theoretically runs several MAPF-LNS independently; however, in comparison to MAPF-LNS, its NPO* and memory usage do not grow in proportion to the number of threads due to the limited memory bandwidth. Along with poor exploitation of the best-known solutions, DETA-LNS results in poor solution quality when the number of threads increases. Still, DETA-LNS can reach higher NPO* and EXP than MAPF-LNS, demonstrating its productivity.

SYNC-LNS waits until all the threads complete their tasks and prunes poor-quality solutions at each iteration. It exploits good-quality solutions and can thus result in higher DP than MAPF-LNS and DETA-LNS in congested instances. At each iteration, if the solution selected by SYNC-LNS during the synchronization always has a lower SOC than the best-known solution, then its EXP becomes $1-\frac{1}{m}$ since the best-known solution is selected from one of the $m$ worker threads. However, since SYNC-LNS requires all threads to wait until each of them finishes its operations, it is less efficient in large maps where finding a path for an agent becomes time-consuming.

DROP-LNS trades off between productivity and synchronization by updating the best-known solution on the fly to improve the solution quality efficiently. Compared to SYNC-LNS, which synchronizes at each iteration, DROP-LNS requires less synchronization overhead, typically resulting in a higher NPO*. Compared to DETA-LNS, which never performs synchronization until the time budget runs out, DROP-LNS has lower AUC since its worker thread exploits the best-known solution synchronized by others. Also, DROP-LNS remains effective in terms of solution quality when the number of threads increases. Compared to LaCAM*, DROP-LNS, even with 8 worker threads, uses significantly less memory and reaches better solution quality and AUC.

\section{Conclusion}
In this paper, we presented DROP-LNS, a parallel framework that performs multiple destroy and repair operations concurrently to explore more regions of the search space within a limited time budget, while the currently best-known solution is updated asynchronously to maintain the productivity of worker threads.
Unlike other parallelism variants, DROP-LNS trades off between productivity and synchronization to reach a better performance overall. It keeps the idle time per worker thread low while still focusing the search on more promising solutions due to the asynchronous updates.
The empirical evaluations confirm our conceptual discussion, showing that DROP-LNS outperforms other parallelism variants and state-of-the-art anytime algorithms such as MAPF-LNS and LaCAM* in six maps from the MAPF benchmark. Future work includes developing more sophisticated mechanisms for synchronization and extensions to anytime bounded-suboptimal algorithms as well as parallel algorithms using GPU.

%%%%%%%%%%%%%%%%%%%%%%%%%%%%%%%%%%%%%%%%%%%%%%%%%%%%%%%%%%%%%%%%%%%%%%%%

\bibliographystyle{named}
\bibliography{arxiv}

\end{document}